\documentclass[prb,preprint, amsmath,amssymb]{revtex4-1}

\usepackage{graphicx}
\usepackage[section]{placeins}
\usepackage{soul}

\usepackage{hyperref}   
\hypersetup{
	colorlinks=true, 
	linkcolor=blue, 
	citecolor=blue, 
	urlcolor=magenta, 
	pdfborder={0 0 0} 
}

\begin{document}
\title{A Magnetic Torsional Pendulum for Exploring Forced Resonance, Parametric Resonance, and Parametric Amplification}

\author{Wenqing Xie}
\thanks{These authors contributed equally to this work.}

\author{Jiahao Wu}
\thanks{These authors contributed equally to this work.}

\author{Yujun Shi}
\email{yujunshi@sxu.edu.cn}
\thanks{Author to whom correspondence should be addressed.}

\affiliation{
	School of Physics and Electronics Engineering,
	Shanxi University, Taiyuan 030006,
	People's Republic of China
}

\begin{abstract}
	We present a magnetic torsional pendulum that provides a unified experimental platform for investigating forced resonance, parametric resonance, and degenerate parametric amplification in the undergraduate laboratory. The system consists of a permanent magnet suspended by thin wires and driven by externally applied magnetic fields generated by Helmholtz coils. By independently controlling a direct driving field and a periodically modulated bias field, the apparatus can realize ordinary forced oscillations, parametric excitation, and phase-sensitive parametric amplification within the same physical system. A miniature wireless gyroscope embedded in the pendulum bob enables direct measurement of the angular velocity and provides convenient real-time acquisition of quantitative dynamical data. A unified equation of motion is derived to describe all three operating regimes. Experimental studies of forced resonance, parametric resonance, and phase-sensitive parametric amplification are compared with theoretical predictions and numerical simulations. The measurements reproduce the characteristic features of all three phenomena and illustrate the influence of nonlinear effects on the system dynamics. The apparatus combines a simple mechanical design, low-cost instrumentation, and highly visible motion. By allowing direct comparison of different resonance mechanisms and their underlying energy-transfer processes,, it provides an accessible platform for studying oscillation theory, nonlinear dynamics, and parametric phenomena in advanced undergraduate laboratories.
\end{abstract}

\maketitle

\section {Introduction}\label{sec1}

In recent years, magnetic pendulum and magnetic oscillator systems based on electromagnetic interactions have attracted increasing attention in undergraduate physics laboratories\cite{10.1119/1.3276412, Wang_2025, Donoso_2012, Egri_2021}. Compared with purely mechanical excitation, electromagnetic actuation provides contactless energy transfer, convenient adjustment of excitation parameters, and precise control of driving conditions. These features make magnetic oscillators particularly suitable for the study of resonance phenomena and nonlinear dynamics.

A particularly simple realization is the magnetic torsional pendulum, in which a permanent magnet suspended by a thin wire interacts with an externally applied magnetic field. Similar systems have been employed to investigate either ordinary forced oscillations\cite{10.1119/1.1969578} or parametric resonance\cite{PhysRevE.56.6613}. In most undergraduate laboratories, however, these two resonance mechanisms are typically introduced through separate experiments\cite{10.1119/1.1969578, PhysRevE.56.6613}. As a result, students often have limited opportunities to directly compare their excitation mechanisms, resonance conditions, and energy-transfer processes within a common physical system.

An interesting connection between the two phenomena emerges when periodic driving and parametric excitation are simultaneously present. Under suitable conditions, the energy supplied through parameter modulation can reinforce an externally excited oscillation, producing parametric amplification\cite{PhysRevLett.67.699, PhysRevE.94.022201, 10.1063/1.3446851, AGHAMOHAMMADI2020103585}. Rather than representing a completely separate resonance phenomenon, parametric amplification can be viewed as the interplay between ordinary resonance and parametric resonance. Consequently, an experimental platform capable of demonstrating all three effects within a single system would provide students with a more unified understanding of resonance phenomena and oscillatory energy transfer.

In this work, we design and construct a magnetic torsional pendulum system that serves as such a unified platform. By controlling the frequency and amplitude of an external magnetic field, the apparatus enables the demonstration and quantitative investigation of forced oscillations, parametric resonance, and parametric amplification using the same experimental setup. To characterize the motion, a miniature wireless gyroscope is embedded inside the pendulum bob, allowing the angular velocity to be measured directly and transmitted in real time. This approach eliminates the need for external optical tracking systems and provides a convenient means of obtaining quantitative dynamical data.

The apparatus features a simple structure, good reproducibility, and highly visible oscillatory motion. Through a series of experiments, students can directly compare different resonance mechanisms and investigate resonance curves, phase relationships, instability thresholds, energy-transfer processes, and nonlinear saturation effects. The experiment therefore provides an accessible and versatile platform for studying resonance phenomena, nonlinear dynamics, and oscillation theory in the undergraduate physics laboratory. 

The remainder of this paper is organized as follows. In Sec.~II, we discuss the operating principles of the magnetic torsional pendulum and derive the governing equation of motion. Section~III describes the experimental apparatus, including the magnetic driving system and the wireless gyroscope used for motion measurements. In Sec.~IV, we present experimental investigations of forced resonance, parametric resonance, and parametric amplification, and compare the characteristics of these oscillatory regimes.  Finally, Sec.~V summarizes the main findings, followed by a discussion on their educational implications and future extensions of this experiment.

\section{Driving Mechanisms and Governing Equation}

\begin{figure}
	\includegraphics[width = 1\textwidth]{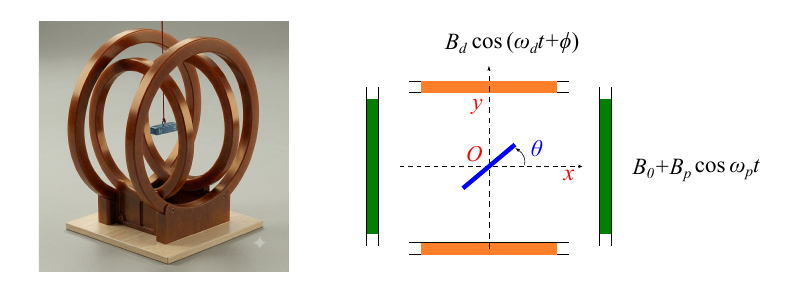}
	\caption{Schematic diagram of the driving Mechanisms. (a) Side view of a horizontally suspended permanent magnet with magnetic moment $m$, positioned at the center of two pairs of Helmholtz coils. (b) Top view of the system showing the relative orientation of the magnet and the applied magnetic fields.}
	
	\label{Schematic diagram}
\end{figure}
\FloatBarrier

\subsection{Unified Governing Equation}

As shown in Fig.~\ref{Schematic diagram}, a permanent magnet with magnetic dipole moment \(m\) is subjected to a periodically time-varying magnetic field \(\mathbf{B}\) generated by two sets of Helmholtz coils. One set produces the direct driving field \(\mathbf{B_d}=B_d \cos \left( \omega_d t + \phi \right)\),  while the other generates both the DC bias field \(\mathbf{B_0}\) and the parametric driving field \(\mathbf{B_p}=B_p \cos \omega_p t\). The total magnetic field in the central region can therefore be written as
\[
\mathbf{B}
=
\left(
B_0 + B_p \cos \omega_p t
\right)\mathbf{e}_x
+
B_d \cos \left( \omega_d t + \phi \right)\mathbf{e}_y .
\]
Let \(\theta\) denote the angular displacement of the magnetic dipole moment from its equilibrium position. The magnetic dipole moment vector can then be written as
\[
\mathbf{m}
=
m\cos\theta\,\mathbf{e}_x
+
m\sin\theta\,\mathbf{e}_y .
\]
The kinetic and potential energies of the torsional pendulum are respectively given by
\begin{equation}
	T=\frac{1}{2}I\dot{\theta}^{2},
\end{equation}

\begin{equation}
	\begin{aligned}
		U
		&=
		\frac{1}{2}K\theta^{2}
		-
		\mathbf{m}\cdot\mathbf{B}
		\\
		&=
		\frac{1}{2}K\theta^{2}
		-
		m\left(
		B_{0}+B_{p}\cos\omega_{p}t
		\right)\cos\theta
		-
		mB_{d}\cos\left(
		\omega_{d}t+\phi
		\right)\sin\theta,
	\end{aligned}
\end{equation}
where \(I\) is the moment of inertia of the torsional pendulum, and \(K\) is the torsional stiffness.

Neglecting damping, the system is conservative and the Lagrangian is given by \(L=T-U\). Applying the Lagrange equation,
\begin{equation}
	\frac{d}{dt}\left(\frac{\partial L}{\partial \dot{\theta}}\right)
	-
	\frac{\partial L}{\partial \theta}
	=0,
\end{equation}
we obtain the equation of motion
\begin{equation}
	I\ddot{\theta}
	+
	\left[
	K + mB_0 + mB_p\cos(\omega_p t)
	\right]\sin\theta
	=
	mB_d\cos(\omega_d t+\phi)\cos\theta .
\end{equation}
Including linear viscous damping, the equation of motion becomes
\begin{equation}
	I\ddot{\theta}
	+ c\dot{\theta} +
	\left[
	K + mB_0 + mB_p\cos(\omega_p t)
	\right]\sin\theta
	=
	mB_d\cos(\omega_d t+\phi)\cos\theta .
\end{equation}
For small angular displacements (\(|\theta|\ll 1\)), we use the approximations
\(\sin\theta \approx \theta\) and \(\cos\theta \approx 1\), yielding
the linearized equation of motion
\begin{equation}
	I\ddot{\theta}
	+c\dot{\theta}
	+\left[
	K+mB_0+mB_p\cos(\omega_p t)
	\right]\theta
	=
	mB_d\cos(\omega_d t+\phi).
\end{equation}

In the present experimental setup, the magnetic restoring torque can be made to dominate over the mechanical torsional stiffness, i.e. \(K \ll mB_0\). In this regime, the contribution of the mechanical torsional stiffness can be neglected. The equation of motion then reduces to
\begin{equation}\label{motion_equation}
	I\ddot{\theta}
	+c\dot{\theta}
	+mB_0\left[
	1+\frac{B_p}{B_0}\cos(\omega_p t)
	\right]\theta
	=
	mB_d\cos(\omega_d t+\phi).
\end{equation}

\subsection{Dimensionless Form}

Let \(\tau=\omega t\), where \( \omega_0{}^2=\frac{mB_0}{I}\). Equation~\ref{motion_equation} can then be nondimensionalized as
\begin{equation}\label{Dimensionless_equation_linear}
	\ddot{\theta}
	+\frac{1}{Q}\dot{\theta}
	+\left[
	1+h\cos\left(\Omega_p\tau\right)
	\right]\theta
	=
	d\cos\left(\Omega_d\tau+\phi\right),
\end{equation}
where
\[
h=\frac{B_p}{B_0},
\qquad
d=\frac{B_d}{B_0},
\qquad
\Omega_p=\frac{\omega_p}{\omega_0},
\qquad
\Omega_d=\frac{\omega_d}{\omega_0},
\]
and \(Q=I\omega_0/c\) is the quality factor. Here, the overdot denotes differentiation with respect to the dimensionless time variable \(\tau\). 

Equation~\ref{Dimensionless_equation_linear} unifies two classical types of oscillatory dynamics. When \(h=0\), it reduces to the conventional damped driven harmonic oscillator. When \(d=0\), it becomes the damped Mathieu equation governing parametric resonance. For angular displacements beyond the small-angle regime, Eq.~\ref{Dimensionless_equation_linear} must be replaced by the full nonlinear equation of motion,
\begin{equation}\label{Dimensionless_equation_nonlinear}
	\ddot{\theta}
	+\frac{1}{Q}\dot{\theta}
	+\left[
	1+h\cos\left(\Omega_p\tau\right)
	\right]\sin\theta
	=
	d\cos\left(\Omega_d\tau+\phi\right)\cos\theta.
\end{equation}

\section{Experimental system}

\subsection{Mechanical setup}
\subsubsection{ Coils}
\begin{figure}[h!]
	\includegraphics[width = 0.85\textwidth]{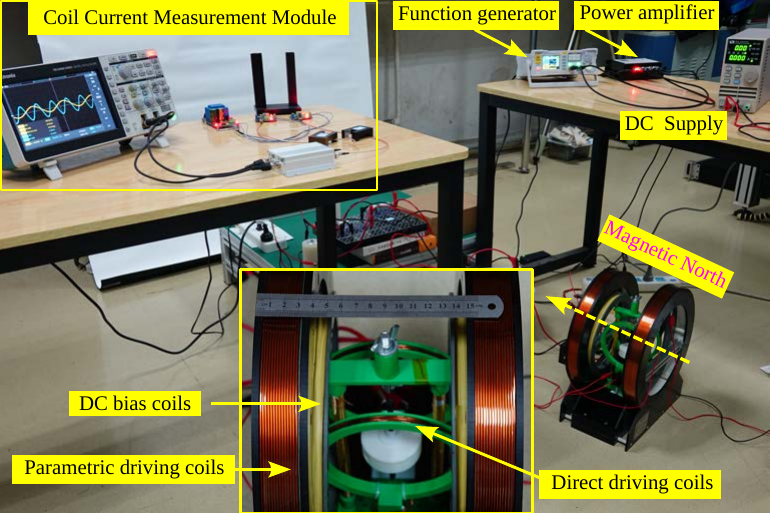}
	\caption{
		Photograph of the complete experimental apparatus.
	}
	
	\label{setup}
\end{figure}

Figure~\ref{setup} shows the complete experimental apparatus, including the coil assembly, the magnetic torsional pendulum, the driving electronics, and the coil current measurement system. Three pairs of Helmholtz coils are used to generate the magnetic fields required for biasing, direct driving, and parametric excitation. The axes of the biasing and parametric-excitation coils are aligned parallel to the horizontal component of the Earth's magnetic field. The axes of the biasing and parametric-excitation coils are aligned parallel to the horizontal component of the Earth's magnetic field, a direction that can be readily identified using a conventional compass or the compass application available on most smartphones. The parameters of the three Helmholtz coil pairs used to achieve the operating conditions of the experiment are summarized in Table~\ref{tab:coils}.

\begin{table}[htbp]
	\centering
	\caption{Parameters of the three Helmholtz coil pairs used in the experiment.}
	\label{tab:coils}
	\begin{tabular}{lcccc}
		\hline
		\textbf{ Coils} &
		\textbf{Equivalent radius} &
		\textbf{Turns per coil} &
		\textbf{Field at center} &
		\textbf{Typical operating field} \\
		&
		(mm) &
		$N$ &
		(G/A) &
		(G) \\
		\hline
		DC biasing         & 105 & 450 & 37.5 & 7.5    \\
		Direct driving     & 60  & 10  & 1.5  & 0.0075 \\
		Parametric driving & 145 & 210 & 10   & 0.1    \\
		\hline
	\end{tabular}
\end{table}

The parametric-excitation coils shown in Fig.~\ref{setup} were adapted from existing laboratory equipment and therefore have a relatively large number of windings. As will be shown in the following sections, the operating conditions required for the present experiment can be achieved with substantially fewer turns. Consequently, the apparatus can be readily reproduced using simpler and more compact coil assemblies.

\subsubsection{ Torsional Pendulum}
\begin{figure}[h!]
	\includegraphics[width = 0.85\textwidth]{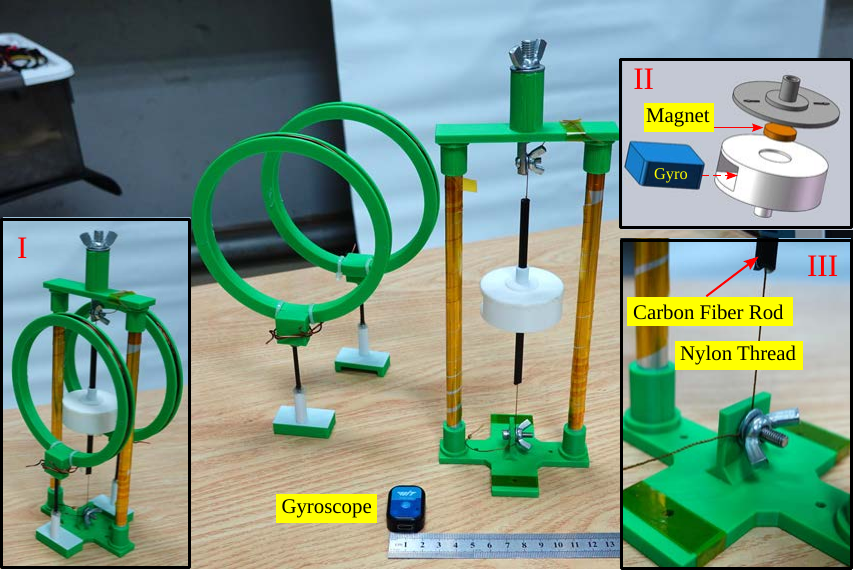}
	
	\caption{
	Magnetic torsional pendulum assembly.
	(I) Assembled torsional pendulum together with the direct-driving coils.
	(II) Schematic illustration showing the embedded permanent magnet and wireless gyroscope inside the pendulum bob.
	(III) Enlarged view of the nylon-thread suspension and mounting structure. 
    }

	\label{pendulum}
\end{figure}

The core of the torsional pendulum is a radially magnetized NdFeB permanent magnet (15~mm in diameter, 3~mm thick, N40 grade) embedded in a 3D-printed circular PLA disk. The disk also contains a compartment for mounting a miniature wireless gyroscope used to measure the angular velocity of the pendulum. The complete pendulum bob, including the embedded gyroscope and two lightweight hollow carbon-fiber tubes used to connect the disk to the suspension thread, has a total mass of approximately 33.8~g, of which the gyroscope accounts for 8.5~g.

Since the experiment is intended to operate in a torsional mode, it is important to suppress unwanted swinging motion. To achieve this, the disk is supported by two thin wires attached to the upper and lower parts of the frame, forming a torsion axis with enhanced mechanical stability. The tension of the upper suspension wire can be adjusted by a screw mechanism, allowing fine alignment of the pendulum and stable operation during long experimental runs.

\subsection{Experimental control and detection}

\begin{figure}[h!]
	\includegraphics[width = 0.75\textwidth]{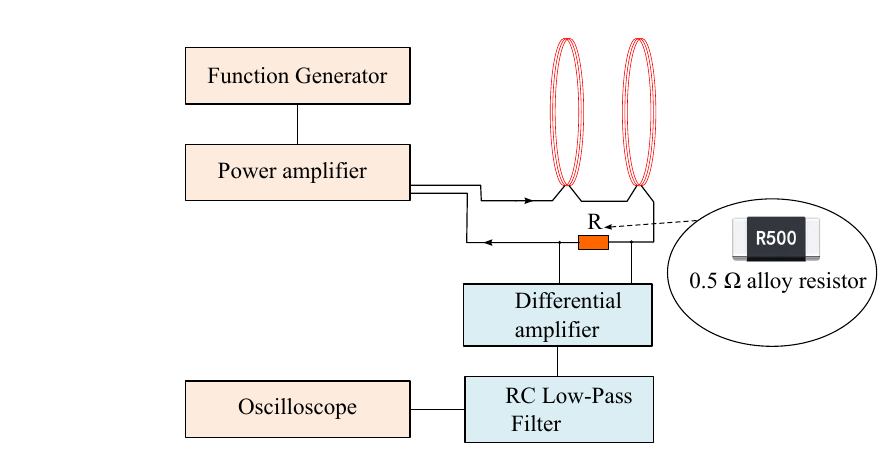}
	\caption{
	Schematic of driving and measuring circuits.
	}
	
	\label{driving}
\end{figure}

\begin{figure}[h!]
	\includegraphics[width = 0.75\textwidth]{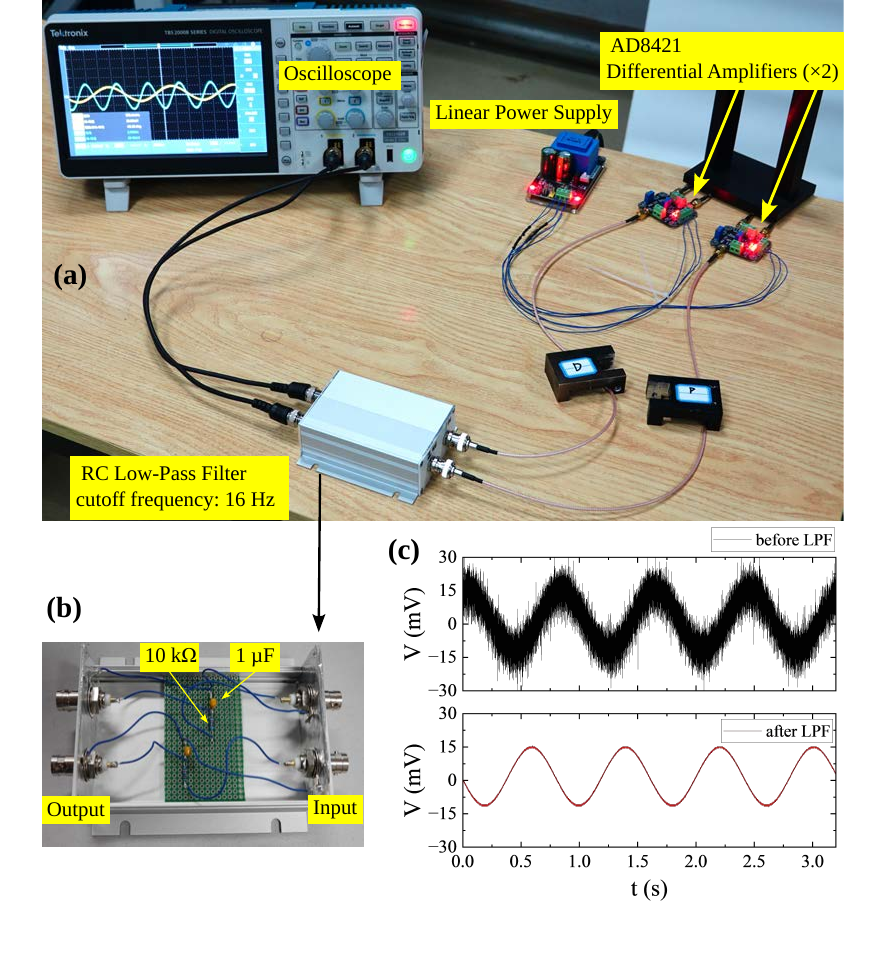}
    \caption{
	Measurement hardware used for monitoring the coil current through the voltage across the current-sensing resistor.
	(a) Photograph of the measurement hardware.
	(b) RC low-pass filter used to suppress high-frequency electrical noise.
	(c) Typical measured signals before and after low-pass filtering. The two traces are sequential measurements rather than simultaneous data; thus, the apparent phase difference is not physically meaningful. The filter improves the signal-to-noise ratio while preserving the relevant low-frequency component.
    }
	
	\label{measuring}
\end{figure}

\subsubsection{Generation and Determination of the Driving Magnetic Field}

As shown in Fig.~\ref{driving}, a dual-channel function generator (FY6900, FeelElec) and a two-channel power amplifier (FPA2000, FeelElec) are used to drive the excitation coils \cite{FeelElec}. Commercial audio amplifiers are generally not ideal for the present application because the operating frequencies of our torsional pendulum are only a few hertz, where the frequency response of many audio amplifiers may deviate significantly from a flat gain profile.

The amplitude of the alternating magnetic field produced by the excitation coils is controlled in an open-loop configuration. The coil current is monitored via the voltage drop across a current-sensing alloy resistor ($R = 0.5~\Omega$) connected in series with each pair of coils. An important experimental consideration is that standard oscilloscope probes measure voltages with respect to the instrument ground and therefore cannot always be connected directly across an arbitrary circuit element. Incorrect probe connections may create unintended current paths and, in some cases, short-circuit the circuit under investigation. Readers unfamiliar with ground-referenced oscilloscope measurements may find the video demonstration in Ref.~\cite{EEVblog279} particularly helpful.  To measure the AC voltage across the sensing resistor, a differential amplifier module (AD8421) is employed. The differential amplifier therefore serves not only to amplify the signal (gain~\(=10\)) but also to perform accurate differential voltage measurements without requiring a dedicated differential oscilloscope probe. The amplified signal is subsequently passed through an RC low-pass circuit before being recorded by the oscilloscope.

Low-pass filtering is particularly useful because the environmental electrical noise introduced by the coils and connecting cables is significant and is primarily concentrated at much higher frequencies (above tens of hertz). The filter consists of a 10~k$\Omega$ resistor and a 1~$\mu$F capacitor, corresponding to a cutoff frequency of approximately 16~Hz. This filter provides substantial attenuation of high-frequency noise while preserving the signal amplitude in the frequency range relevant to the experiment. Figure~\ref{measuring} shows the measurement hardware. A comparison of the signals before and after filtering demonstrates a significant improvement in the signal-to-noise ratio.

\subsubsection{Angular Velocity Measurement Using a Wireless Gyroscope}

The torsional oscillation is characterized by measuring the angular velocity using a miniature wireless gyroscope (WT901BC, WitMotion Shenzhen Co., Ltd.)\cite{WitMotion_WT901BC}, sampled at 50 Hz. Compared with video-based tracking methods, the gyroscope provides higher temporal resolution, simpler data acquisition, and lower cost for consumer-grade applications.

\section{Examples of results with the pendulum setup}

\subsection{Free oscillations }\label{Free oscillations}
\begin{figure}
	\includegraphics[width = 1\textwidth]{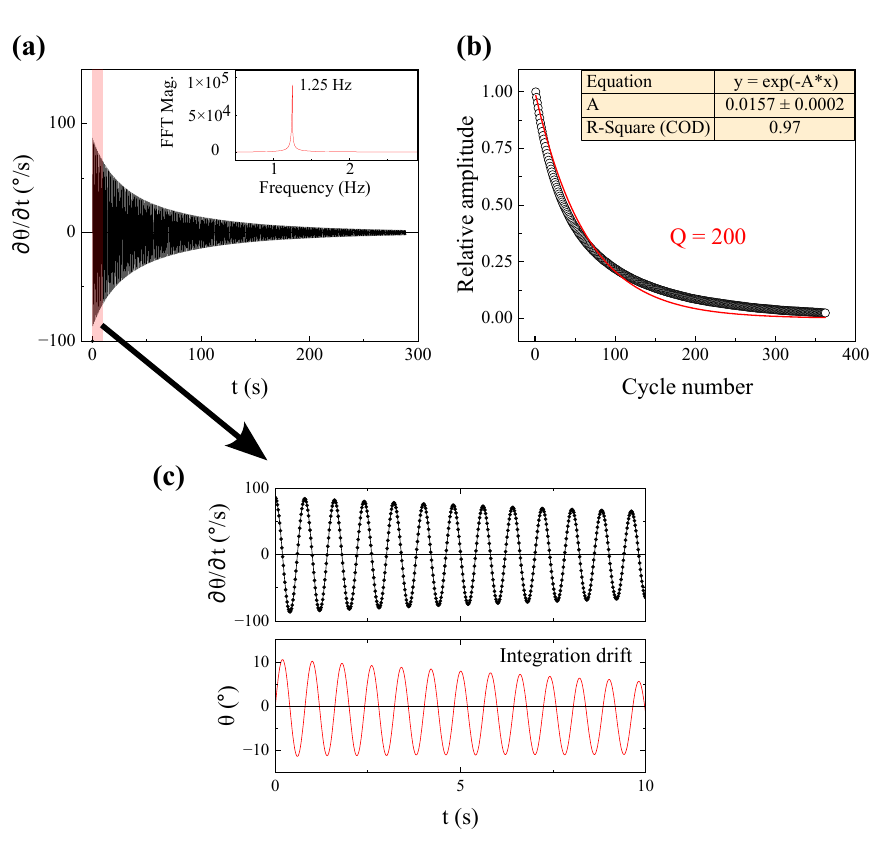}
	\caption{Free oscillation of the torsional pendulum (in physical time units). 
		(a) Angular velocity as a function of time; the inset shows its FFT spectrum. 
		(b) Relative amplitude as a function of the number of oscillation cycles. 
		(c) Zoomed-in view of the first 10~s of the angular velocity shown in panel (a), together with the angular displacement obtained by numerical integration of the angular velocity signal.
		}
	\label{Free_oscillation_01}
\end{figure}

The free oscillation is well described by a linearly damped harmonic oscillator,
\begin{equation}\label{Eq_damped_free_oscillation}
	\ddot{\theta} + \frac{1}{Q}\dot{\theta} + \theta = 0,
\end{equation}
which corresponds to Eq.~\ref{Dimensionless_equation_linear} in the absence of parametric modulation ($h=0$) and external driving ($d=0$). In the underdamped regime ($Q > 1/2$), the solution exhibits an exponentially decaying envelope. For $Q \gg 1$, the oscillation frequency is approximately unity, and the envelope of the angular velocity (and displacement) decays as $e^{-\tau/(2Q)}$. Sampling this envelope at successive oscillation periods yields
\begin{equation}
	|\dot{\theta}|_n \propto e^{-\frac{\pi}{Q} n},
\end{equation}
where $n$ denotes the oscillation cycle number.

Figure~\ref{Free_oscillation_01} shows the free oscillation of the torsional pendulum in the presence of a DC bias magnetic field ($7.5~\mathrm{G}$), with both the driving and parametric-excitation fields turned off. The natural frequency is $f_0 = 1.25~\mathrm{Hz}$, corresponding to $\omega_0 = 2\pi f_0$, and the quality factor is $Q \sim 200$, obtained from fitting the decay envelope shown in Fig.~\ref{Free_oscillation_01}(b). 
Figure~\ref{Free_oscillation_01}(c) shows a zoomed-in view of the first 10 s of the motion shown in panel (a). The angular displacement, reconstructed by numerical integration of the measured angular velocity, exhibits a noticeable integration drift, which arises from the static bias drift of the MEMS gyroscope on the order of $(0.5\,\text{--}\,1)^{\circ}\!/\mathrm{s}$

\begin{figure}
	\includegraphics[width = 1\textwidth]{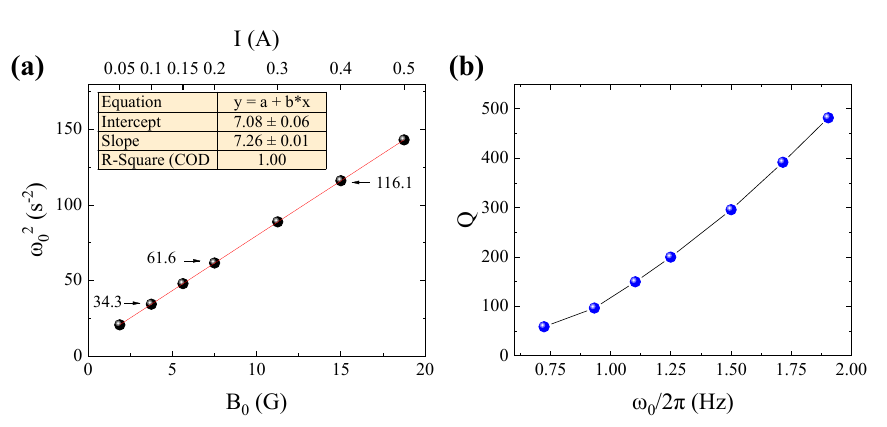}
    \caption{
	(a) Natural frequency as a function of the DC bias magnetic field. The upper x-axis shows the corresponding coil current.
	(b) Quality factor as a function of the natural frequency.
    }
	\label{Free_oscillation_02}
\end{figure}

Figure~\ref{Free_oscillation_02} shows the free oscillation of the torsional pendulum for different values of the DC bias magnetic field. In panel (a), a linear dependence between $\omega_0^2$ and $B_0$ is observed. The fitted line passes close to the origin, with a small intercept, which supports the validity of the model described by Eq.~\ref{motion_equation} in the absence of external driving. Panel (b) shows the dependence of the quality factor on  the natural frequency. Within the investigated range (0.72–1.9~Hz), the quality factor increases with frequency.

The reported values correspond to a global quality factor $Q_{\mathrm{global}}$ obtained from fitting the entire decay envelope. A local quality factor $Q_{\mathrm{local}}$ can be defined by restricting the fit to a short time window, which is found to increase as the oscillation amplitude decays (see Appendix~A). This behavior suggests that the quality factor is not a constant, but rather reflects an amplitude-dependent effective dissipation in the system. A more accurate description would require going beyond the linear damping approximation; however, this effect is beyond the scope of the present work.

\FloatBarrier

\subsection{Forced oscillation}

The forced oscillation is described by
\begin{equation}\label{Eq_direct_driving_linear}
	\ddot{\theta}
	+\frac{1}{Q}\dot{\theta}
	+\theta
	=
	d\cos\left(\Omega_d\tau+\phi\right),
\end{equation}
within the small-angle approximation, or more accurately by
\begin{equation}\label{Eq_direct_driving_nonlinear}
	\ddot{\theta}
	+\frac{1}{Q}\dot{\theta}
	+\sin\theta
	=
	d\cos\left(\Omega_d\tau+\phi\right)\cos\theta.
\end{equation}
These equations correspond to Eqs.~\ref{Dimensionless_equation_linear} and \ref{Dimensionless_equation_nonlinear}, respectively, in the absence of parametric modulation ($h=0$). 

\begin{figure}
	\includegraphics[width = 1\textwidth]{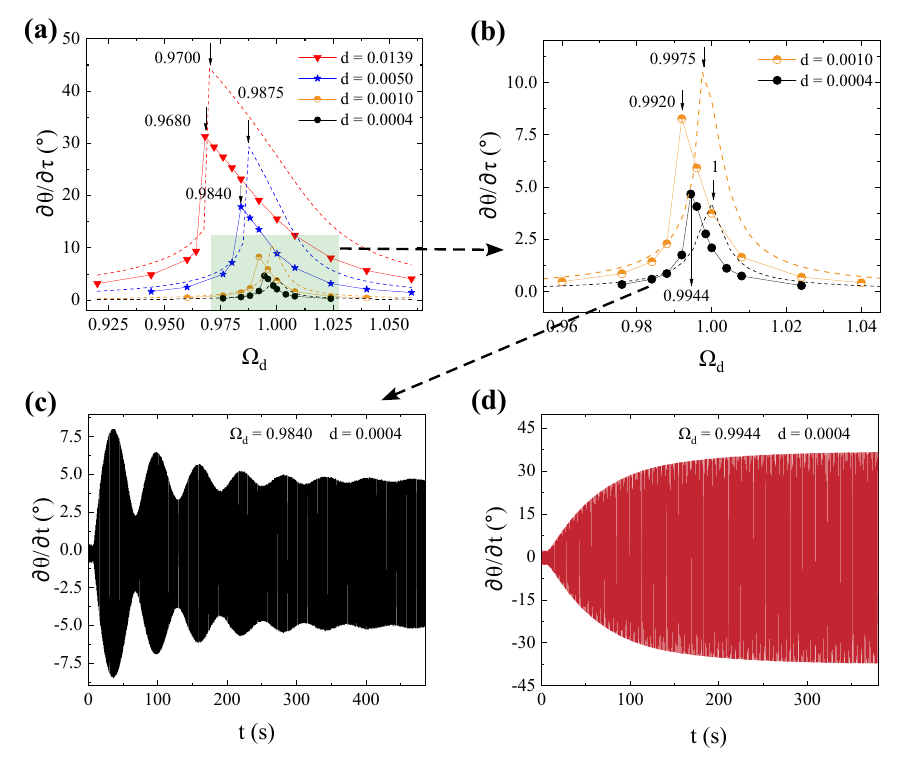}
   \caption{
   	Experimental results of the forced oscillation.
   	(a--b) Steady-state frequency response of the angular velocity amplitude (dimensionless) as a function of driving frequency for different driving strengths $d$. The dashed curves are obtained from numerical solutions of Eq.~\ref{Eq_direct_driving_nonlinear}.
   	(c--d) Time traces of the system response (in physical units) below resonance ($\Omega_d=0.9840$) and near resonance ($\Omega_d=0.9944$), respectively, for a driving strength $d=0.0004$.
   }
	\label{Forced oscillation}
\end{figure}

Figure~\ref{Forced oscillation} shows typical experimental results for forced oscillations, with a natural frequency of $1.25~\mathrm{Hz}$ and a quality factor of $189$, under a DC bias field of 7.5~G. The slightly different value of the quality factor compared with that in Fig.~\ref{Free_oscillation_01} ($Q \approx 200$) arises because the two measurements were not performed within the same experimental run. In particular, each insertion of the wireless gyro introduces a small perturbation to the torsional pendulum parameters. Therefore, within each full set of forced-oscillation measurements and subsequent vibration experiments, the gyro is kept fixed in place to avoid such run-to-run variations.

Figure~\ref{Forced oscillation}(a) and (b) show the steady-state frequency response of the angular velocity amplitude as a function of driving frequency for different driving strengths $d$. For each scanning point, the system is initialized close to rest, with the measured angular velocity below $1^\circ/\mathrm{s}$. With increasing oscillation amplitude, nonlinear effects in the torsional pendulum become more pronounced, resulting in a softening of the resonance frequency, consistent with the prediction of  Eq.~\ref{Eq_direct_driving_nonlinear}. Figure~\ref{Forced oscillation}(c) and (d) show the  Time traces of the system response (in physical units) below resonance ($\Omega_d=0.9840$) and near resonance ($\Omega_d=0.9944$), respectively, for a driving strength $d=0.0004$. This driving strength corresponds to a coil-generated magnetic field of approximately $0.003\,\mathrm{G}$, which is much smaller than the typical magnitude of the Earth's magnetic field ($\sim 0.5\,\mathrm{G}$). Nevertheless, a pronounced resonant response is observed, with the angular displacement amplitude reaching approximately $5^\circ$ in Fig.~\ref{Forced oscillation}(d).

\FloatBarrier

\subsection{Parametric Resonance}

The parametric oscillation is described by
\begin{equation}\label{Dimensionless_mathieu_linear}
		\ddot{\theta}
		+\frac{1}{Q}\dot{\theta}
		+\left[
		1+h\cos\left(\Omega_p\tau\right)
		\right]\theta
		=
		0,
\end{equation}
within the small-angle approximation, or more accurately by
\begin{equation}\label{Dimensionless_mathieu_nonlinear}
	\ddot{\theta}
	+\frac{1}{Q}\dot{\theta}
	+\left[
	1+h\cos\left(\Omega_p\tau\right)
	\right]\sin\theta
	=
	0.
\end{equation}
These equations correspond to Eqs.~\ref{Dimensionless_equation_linear} and \ref{Dimensionless_equation_nonlinear}, respectively, in the absence of driving forcing ($d=0$). 

Using the linear damped Mathieu equation, Eq.~\ref{Dimensionless_mathieu_linear}, as an example, the oscillation amplitude is predicted to grow exponentially when the excitation frequency and modulation depth fall within certain instability regions, commonly known as Arnold tongues. The principal instability region is centered around $\Omega_p = 2\omega_0$, while higher-order instability regions occur near $\Omega_p = 2\omega_0/n$ ($n=2,3,\ldots$). For the principal resonance, the onset of parametric instability is predicted when the modulation depth \(h\) exceeds the threshold value $h_{\mathrm{th}}=2/Q$\cite{10.1115/1.4039144}. In a real physical system described by Eq.~\ref{Dimensionless_mathieu_nonlinear}, however, the oscillation amplitude cannot increase indefinitely. As the amplitude grows, nonlinear effects become increasingly important and modify the dynamics. As a result, the exponential growth predicted by the linear theory is gradually suppressed, and the system eventually approaches a finite steady-state amplitude.

\begin{figure}
	\includegraphics[width = 1\textwidth]{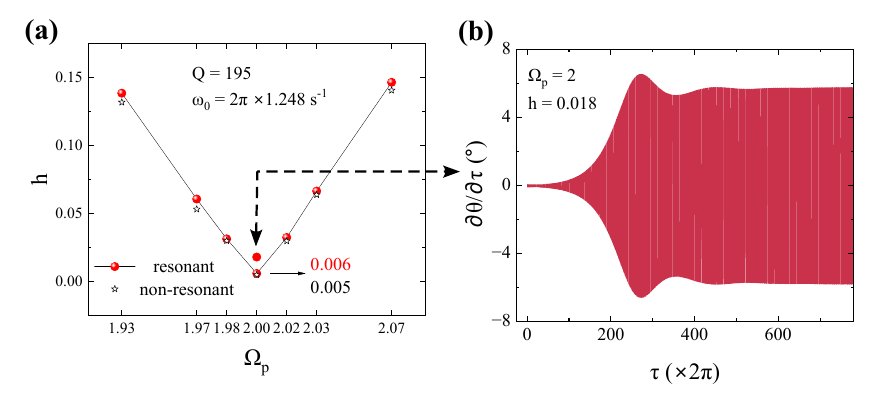}
    \caption{
	Experimental results of parametric oscillation.
	(a) Instability chart of the parametric resonance. Solid circles and open stars denote the presence and absence of parametric resonance, respectively. The solid line serves only as a guide to the eye, indicating the approximate boundary between the unstable and stable regions.
	(b) Time evolution of the dimensionless angular velocity amplitude under parametric excitation.
    }
	\label{Parametric_oscillation}
\end{figure}

Figure~\ref{Parametric_oscillation} shows typical experimental results for parametric oscillations obtained with a natural frequency of $1.248~\mathrm{Hz}$ and a quality factor of $195$. For excitation near the principal parametric resonance ($\Omega_p = 2\omega_0$), the experimentally observed threshold modulation depth is approximately $h_{\mathrm{th}}\approx0.006$, corresponding to a modulation field amplitude of about $0.045~\mathrm{G}$ (i.e., $0.006\times7.5~\mathrm{G}$). It is smaller than the value $2/Q\approx0.010$ predicted using the measured quality factor. This discrepancy can be attributed to the amplitude dependence of the damping. The quality factor $Q=195$ here represents an effective value determined from the overall decay process. At the onset of parametric oscillation, however, the oscillation amplitude is very small and the corresponding damping is weaker, resulting in a larger effective quality factor than the measured global value. Consequently, the instability threshold is reduced. 

\FloatBarrier

\subsection{Parametric Amplification}

In this section, we focus on degenerate parametric amplification (DPA), for which the pump frequency (parametric excitation) satisfies $\Omega_p=2$ and the signal frequency (direct excitation) satisfies $\Omega_d=1$, corresponding to $\Omega_d=\Omega_p/2$ in Eqs.~\ref{Dimensionless_equation_linear} and \ref{Dimensionless_equation_nonlinear}. For brevity, we do not reproduce the theoretical analysis or the explicit gain expressions\cite{PhysRevLett.67.699, 10.1063/1.3446851,AGHAMOHAMMADI2020103585} and instead summarize the main results needed for interpreting the experimental observations.

Let the gain of the parametric amplifier be defined as

\[
G(\phi)=
\frac{\left|A\right|_{\mathrm{pump\,on}}}
{\left|A\right|_{\mathrm{pump\,off}}},
\]
where $\left|A\right|$ denotes the steady-state response amplitude of the system, expressed either in terms of angular displacement or angular velocity.

\begin{figure}
	\includegraphics[width = 1\textwidth]{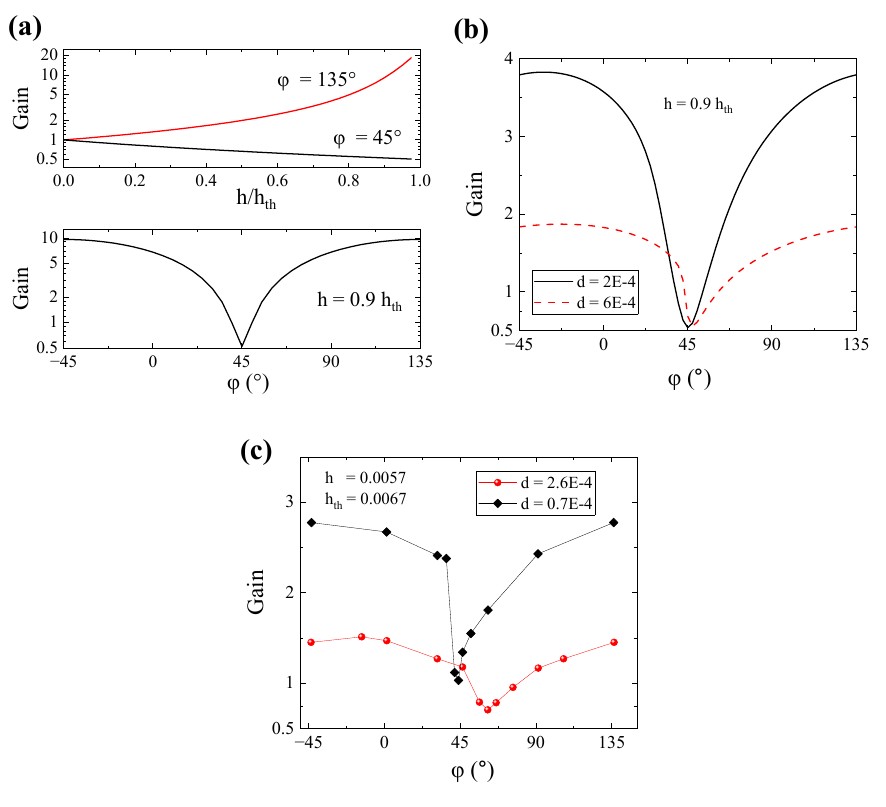}
    \caption{
	Degenerate parametric amplification (DPA).
	(a) Numerical results for the linear model, Eq.~\ref{Dimensionless_equation_linear}. The lower panel shows the gain as a function of the relative phase between the signal and the pump, while the upper panel shows the maximum and minimum gain as functions of the parametric modulation depth $h$.
	(b) Phase dependence of the gain obtained from numerical solutions of the nonlinear model, Eq.~\ref{Dimensionless_equation_nonlinear}, for different signal amplitudes $d$.
	(c) Experimental measurements of the phase-dependent gain of the magnetic torsional pendulum for different signal amplitudes. The phase difference is directly extracted from the oscilloscope-measured  waveforms of the two driving channels, after correcting for systematic phase delays introduced by the driving and measurement electronics.
	}
	\label{DPA}
\end{figure}

For the linear system described by Eq.~\ref{Dimensionless_equation_linear}, the amplifier is operated below the parametric instability threshold ($h<2/Q$). In this regime, the amplification is phase sensitive: the gain depends strongly on the relative phase $\phi$ between the signal and the pump. The gain is minimized at $\phi=\pi/4$ and maximized at $\phi=3\pi/4$, as shown in Fig.~\ref{DPA}(a). For the nonlinear system described by Eq.~\ref{Dimensionless_equation_nonlinear}, the analysis becomes considerably more involved. Nevertheless, phase-sensitive parametric amplification remains clearly observable. Compared with the linear case, nonlinear effects reduce the amplifier performance, leading to a lower gain and a distortion of the phase-dependent response, as shown in Fig.~\ref{DPA}(b). The deviation from the ideal linear behavior becomes increasingly pronounced as the amplitude of the direct excitation (signal) is increased. For a fixed pump strength, a larger signal produces a larger oscillation amplitude, causing the system to explore regions where the nonlinear terms in Eq.~\ref{Dimensionless_equation_nonlinear} become significant. This behavior is consistent with the origin of the nonlinearity, since the restoring and driving torques contain the amplitude-dependent terms $\sin\theta$ and $\cos\theta$, whose departures from their small-angle approximations grow with increasing oscillation angle.

Figure~\ref{DPA}(c) presents the experimental results of the degenerate parametric amplifier (DPA) realized with the magnetic torsion pendulum. In the experimental implementation, the phase difference used in the analysis is obtained from oscilloscope measurements of the two driving signals, with systematic phase delays introduced by the driving and measurement electronics—such as the power amplifier, differential amplifier, and RC filtering circuit—corrected. Although the pump and signal amplitudes used in Figs.~\ref{DPA}(b) and \ref{DPA}(c) are not exactly identical, the two panels allow a qualitative comparison between the numerical predictions and the experimental observations.
The measured gain exhibits a clear dependence on the relative phase between the signal and the pump, confirming the phase-sensitive nature of the parametric amplification process. The overall shape of the gain curve and its evolution with increasing signal amplitude are reproduced reasonably well by the nonlinear model. Nevertheless, noticeable discrepancies remain in the vicinity of the gain minimum. In particular, for weak signal amplitudes, the minimum measured gain remains close to unity rather than approaching the value of $0.5$ predicted by the numerical simulations. As the signal amplitude increases, the phase corresponding to the minimum gain also exhibits a larger shift than that predicted by the model. These differences suggest that additional effects not included in Eq.~\ref{Dimensionless_equation_nonlinear}.

A possible origin of these discrepancies is the amplitude-dependent damping discussed in Sec.~\ref{Free oscillations}. The dominant nonlinear terms included in Eq.~\ref{Dimensionless_equation_nonlinear} originate from the trigonometric functions $\sin\theta$ and $\cos\theta$, and therefore become progressively weaker as the oscillation amplitude decreases. In contrast, the measured damping exhibits the opposite trend: the quality factor is not constant but increases significantly as the oscillation amplitude becomes smaller. Consequently, reducing the oscillation amplitude does not necessarily bring the system closer to the ideal harmonic-oscillator model assumed in the simulations. While this effect produces only minor corrections in conventional forced oscillations and parametric resonance experiments, it may become more pronounced in degenerate parametric amplification, where the gain results from a coherent interference process that is highly sensitive to small perturbations of the oscillator dynamics. Such sensitivity may account for the observed deviations near the deamplification condition. Whether amplitude-dependent damping is indeed responsible for these deviations remains an open question and merits further investigation.

\FloatBarrier

\section{Discussion and Conclusion }

We have designed and constructed a magnetic torsional pendulum that provides a unified experimental platform for demonstrating and quantitatively investigating forced resonance, parametric resonance, and parametric amplification. This experiment allows students to connect theoretical models with real dynamical behavior, observe resonance and instability phenomena in real time, and build physical intuition for nonlinear oscillatory systems. In addition, students can further gain hands-on experience with data acquisition, frequency-response analysis, and the role of dissipation and nonlinearity in oscillatory systems.

Like most macroscopic mechanical pendulums, the present system operates at frequencies of only a few hertz. This low-frequency regime makes the motion readily observable and allows students to follow the system dynamics in real time. Furthermore, the transient-to-steady-state evolution typically unfolds over several minutes, providing a clear illustration of the distinction between transient and steady-state behavior. The low operating frequency and relatively high quality factor ($Q \sim \text{several hundred}$) lead to a long relaxation time, making the acquisition of a complete dataset time-consuming. For example, obtaining a single steady-state data point typically requires approximately 10 minutes. This limitation could be alleviated by employing higher-frequency oscillatory systems, such as vibrating strings\cite{PhysRevLett.117.214101}. A related experimental approach was reported in Ref.~\cite{PhysRevLett.117.214101}, where a guitar string was used to demonstrate several phenomena associated with forced and parametric oscillations. Owing to the higher operating frequency and shorter relaxation time, the measurements can be performed more efficiently than in the present system. Interested readers may wish to explore similar experiments as a natural extension of the work presented here.

\section*{Acknowledgments}
This work was supported by the 2026 Teaching Instrument Development Project of the School of Physics and Electronic Engineering at Shanxi University.

\section*{Author Declarations}
\textbf{Conflict of Interest}

The authors have no conflicts of interest to disclose.

\appendix  
\renewcommand{\theequation}{A\arabic{equation}}
\setcounter{equation}{0}

\section*{Appendix: Quality Factor versus Amplitude}\label{Appendix}

\begin{figure}
	\includegraphics[width = 1\textwidth]{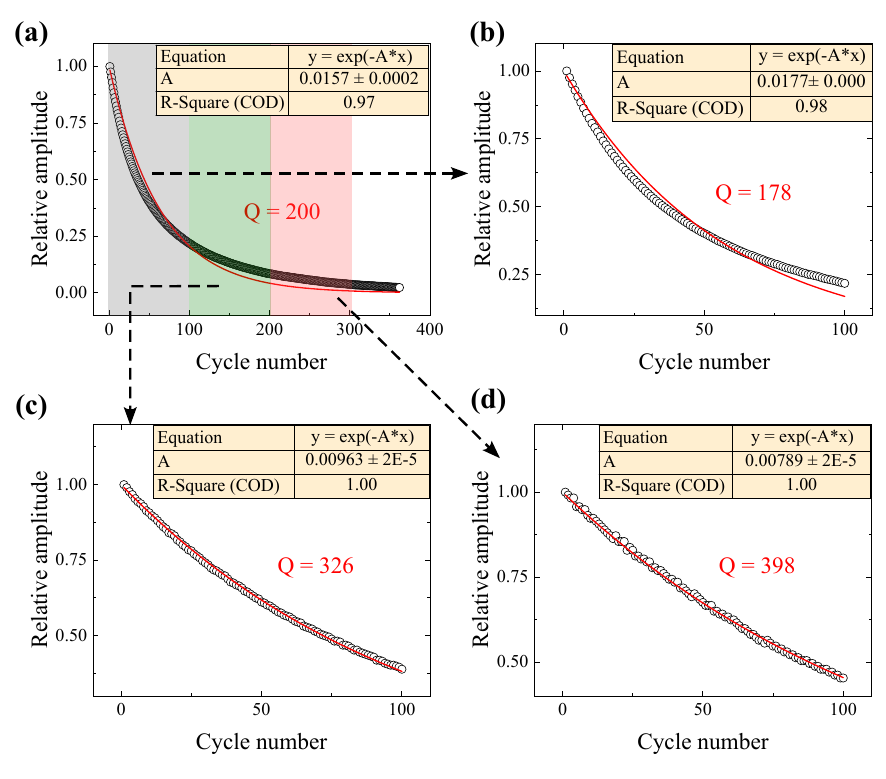}
	\caption{
		 The dependence of the quality factor on the oscillation amplitude.
	}
	\label{Q_A}
\end{figure}

Figure~\ref{Q_A} shows the dependence of the quality factor on the oscillation amplitude. The complete decay curve shown in Fig.~\ref{Q_A}(a) (identical to Fig.~\ref{Free_oscillation_01}(b)) was divided into three consecutive time intervals. Each segment was normalized and fitted independently, yielding a corresponding local quality factor. As shown in Fig.~\ref{Q_A}(b--d), the quality factor is not constant but increases significantly as the oscillation amplitude decreases. This trend is consistently observed throughout our measurements and cannot be attributed to random fluctuations in a particular data set. Instead, it indicates the presence of an amplitude-dependent dissipation mechanism, for which the effective damping becomes weaker at smaller oscillation amplitudes.

\FloatBarrier

\bibliographystyle{unsrt}
\bibliography{reference.bib}

\end{document}